\documentclass[aps,twocolumn,epsfig,superscriptaddress,showpacs]{revtex4}
\usepackage{graphicx} 
\usepackage{psfrag} 
\begin{document}

\title{Insights into exfoliation possibility of MAX phases to MXenes}

\author{Mohammad Khazaei}
\email{khazaei@riken.jp}
\affiliation{Computational Materials Science Research Team, RIKEN Advanced Institute for Computational Science (AICS), Kobe, Hyogo 650-0047, Japan}
\author{Ahmad Ranjbar}
\affiliation{Computational Materials Science Research Team, RIKEN Advanced Institute for Computational Science (AICS), Kobe, Hyogo 650-0047, Japan}
\author{Keivan Esfarjani}
\affiliation{Department of Mechanical and Aerospace Engineering, University of Virginia, 122 Engineer's Way, Charlottesville,VA 22904, USA}
\author{Dimitri Bogdanovski}
\affiliation{Chair of Solid State and Quantum Chemistry, RWTH Aachen University, 52056 Aachen, Germany}
\author{Richard Dronskowski}
\affiliation{Chair of Solid State and Quantum Chemistry, RWTH Aachen University, 52056 Aachen, Germany}
\author{Seiji Yunoki}
\affiliation{Computational Materials Science Research Team, RIKEN Advanced Institute for Computational Science (AICS), Kobe, Hyogo 650-0047, Japan}
\affiliation{Computational Condensed Matter Physics Laboratory, RIKEN, Wako, Saitama 351-0198, Japan}
\affiliation{Computational Quantum Matter Research Team, RIKEN Center for Emergent Matter Science (CEMS), Wako, Saitama 351-0198, Japan}

\date{\today}

\begin{abstract}
Chemical exfoliation of MAX phases into two-dimensional (2D) MXenes can be considered as a major breakthrough in the
synthesis of novel 2D systems. To gain insight into the exfoliation possibility of MAX phases and to identify which 
MAX phases are promising candidates for successful exfoliation into 2D MXenes, we perform extensive electronic structure 
and phonon calculations, and determine the force constants, bond strengths, and static exfoliation energies of 
MAX phases to MXenes for 82 different experimentally synthesized crystalline MAX phases. Our results show 
a clear correlation between the force constants and the bond strengths. As the total force constant of an ``A'' 
atom contributed from the neighboring atoms is smaller, the exfoliation energy becomes 
smaller, thus making 
exfoliation easier. We propose 37 MAX phases for successful exfoliation into 2D Ti$_2$C, Ti$_3$C$_2$, Ti$_4$C$_3, $Ti$_5$C$_4$, 
Ti$_2$N, Zr$_2$C, Hf$_2$C, V$_2$C, V$_3$C$_2$, V$_4$C$_3$, Nb$_2$C, Nb$_5$C$_4$, Ta$_2$C, Ta$_5$C$_4$,
Cr$_2$C, Cr$_2$N, and Mo$_2$C MXenes. In addition, we explore 
the effect of charge injection on MAX phases. We find that the injected charges, both electrons and holes, are 
mainly received by the transition metals. This is due to the electronic property of MAX phases that the states near 
the Fermi energy are mainly dominated by $d$ orbitals of the transition metals. For negatively charged MAX phases, 
the electrons injected cause swelling of the structure and elongation of the bond distances along the $c$ axis, 
which hence weakens the binding. For positively charged MAX phases, on the other hand, the bonds become 
shorter and stronger. Therefore, we predict that 
the electron injection by electrochemistry or gating techniques can significantly facilitate the exfoliation 
possibility of MAX phases to 2D MXenes.
\end{abstract}

\maketitle

\section{INTRODUCTION}
\begin{figure*}[t]
\centering
  \includegraphics[scale=0.35]{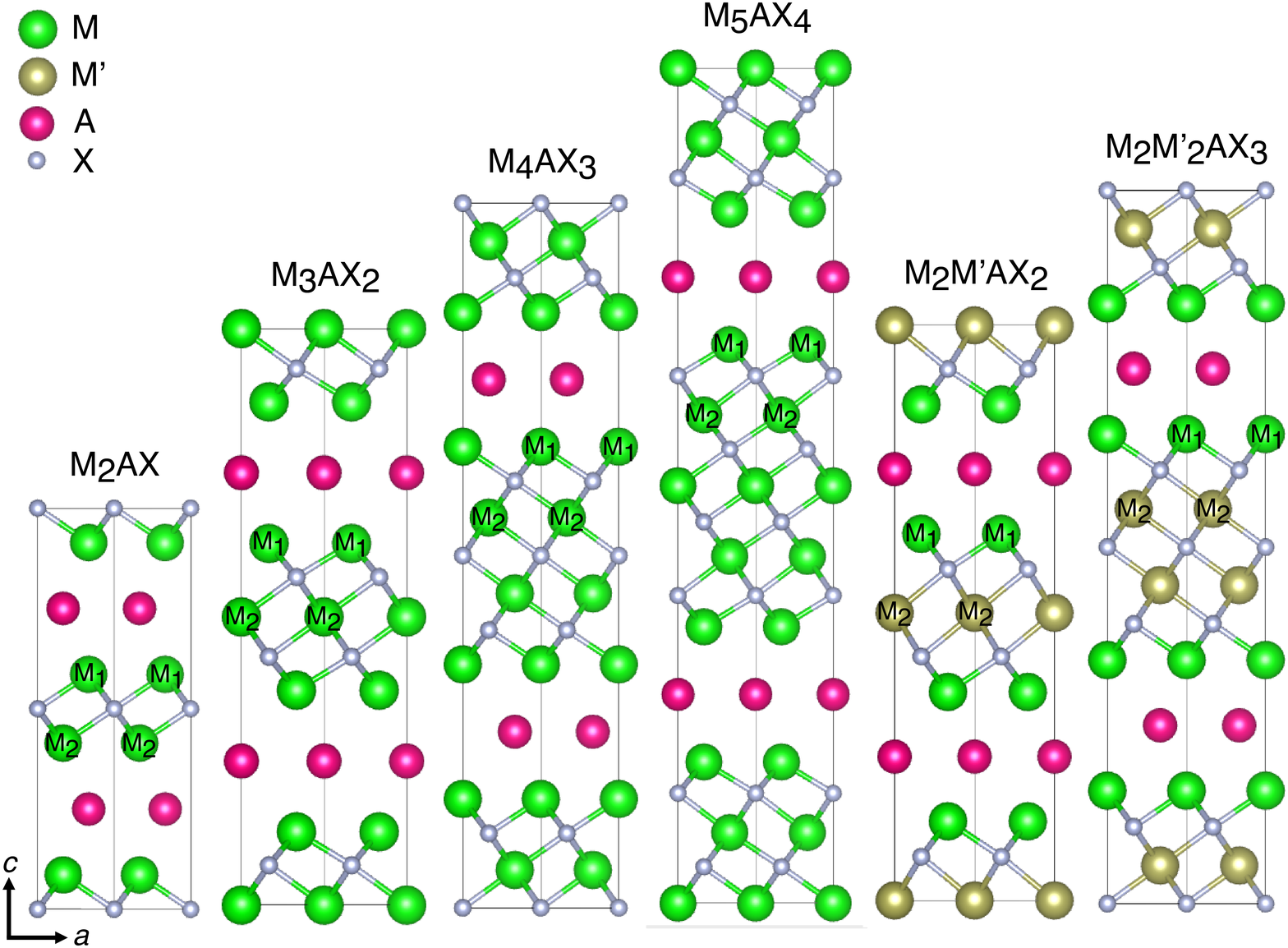}
  \caption{
  Atomic structures of six different crystalline MAX phases. The $a$ and $c$ axes are common to all structures. 
Transition metals in different layers are indicated by ``M$_1$'' and ``M$_2$'' to 
denote different bonds between transition metals 
located in the same layer (M$_1-$M$_1$ or M$_2-$M$_2$ ) and in different layers (M$_1-$M$_2$). 
In M$_{n+1}$AX$_n$ ($n=1-4$),
M$_1$ and M$_2$ are the same type of transition metal, while they are different types in M$_2$M$'$AX$_2$ 
and M$_2$M$'_2$AX$_3$.} 
  \label{fig:maxphases}
\end{figure*}

MAX phases are a large family of solid layered transition metal carbides or nitrides with a hexagonal lattice 
(space group $P6_3/mmc$) and the chemical formula of 
M$_{n+1}$AX$_n$, where ``M'' stands for an early transition metal (Sc, Ti, Zr, Hf, V, Nb, Ta, Cr, or Mo), 
``A'' represents an element from main groups III$-$VI of the periodic table
(Al, Ga, In, Tl, Si, Ge, Sn, Pb, P, As, Bi, S, or Te), ``X'' is carbon or nitrogen, and $n$ = 1$-$4.\cite{M.W.Barsoum2000,J.Wang2009,Z.M.Sun2011,H.Fashandi2017,D.Horlait2016} 
Recently, the family of crystalline MAX phases has been expanded even further. 
For example, a set of ordered double transition metals MAX phases of M$_2$M$'$AX$_2$, 
M$_2$M$'_2$AX$_3$,\cite{B.Anasori2015_1,R.Meshkian2017} 
and (M$_{2/3}$M$'_{1/3}$)$_2$AX$_2$ has been 
synthesized,\cite{Q.Tao2016,M.Dahlqvist2017} where ``M$'$'' is another transition metal. 
The crystal structures of MAX phases with varying stoichiometry are shown schematically in Fig.~\ref{fig:maxphases}. 
To the best of our knowledge, 82 different crystalline 
MAX phases have been experimentally fabricated~\cite{M.W.Barsoum2000,Z.M.Sun2011} as listed in Table 1S. 
In addition to the crystalline MAX phases, various alloys of MAX phases with 
different mixtures of transition metals and/or X elements have also been synthesized. Theoretically, thousands of crystalline or alloy structures of MAX phases have been 
predicted to be stable.\cite{A.Talapatra2016,M.Ashton2016_1,M.Dahlqvist2010,M.Dahlqvist2015,Y.Mo2012,S.Aryal2014,M.F.Cover2009,M.Khazaei2014_1,M.Khazaei2014_2,R.Arroyave2017,C.Jiang2017}

All MAX phases are metallic~\cite{M.Khazaei2014_1,M.Khazaei2014_2,Z.Sun2005,M.A.Ghebouli2011,M.T.Nasir2014,M.A.Hadi2016,M.A.Hadi2017} because of the existence of partially occupied $d$ orbitals of transition metals 
near the Fermi energy.\cite{M.Khazaei2014_1,M.Khazaei2014_2} In other words, due to the metallic bonding between the transition metal atoms either on the same layer or across layers, 
MAX phases become electrically conductive. 
Since MAX phases are made of metal and nonmetal elements and are synthesized at high temperatures, they are usually considered as ceramics. Owing to their high mechanical and thermal 
stability,\cite{S.Aryal2014,M.F.Cover2009,M.Khazaei2014_1,M.Khazaei2014_2,Z.Sun2004_a,J.D.Hettinger2005} some of the MAX phases could be used in harsh conditions, \textit{i.e.}, 
in conditions where the materials should be resistant to thermal shock, fatigue, creep, oxidation, or corrosive reactions.\cite{M.W.Barsoum2000,J.Wang2009}
To enhance the technological applications of MAX phases, the ductility of MAX phases can be improved 
by lowering their ceramic brittleness without harming their toughness when the number of valence 
electrons per unit cell is properly controlled by mixing different M, A, or X elements.\cite{M.F.Cover2009,M.Khazaei2014_1}

\begin{figure*}[t]
\centering
 \includegraphics[scale=0.35]{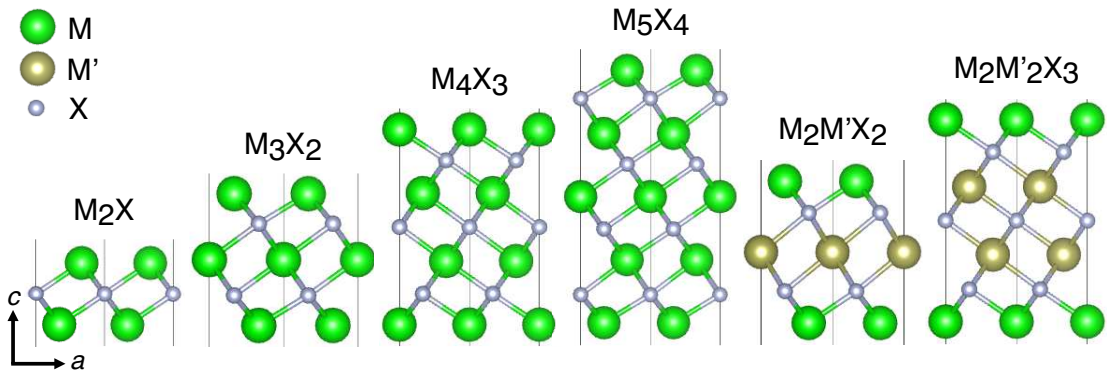}
 \caption{
 Atomic structures of six different 2D MXenes. The $a$ and $c$ axes are common to all structures.
 }
 \label{fig:maxenes}
\end{figure*}

Up to a few years ago, the application of MAX phases was based solely on their interesting mechanical properties, 
rather than their electronic, magnetic, or optical ones. However, the possibility of exfoliation from MAX phases 
into two-dimensional (2D) transition metal carbides or nitrides, so called MXenes, 
has significantly broadened the range of potential applications of MAX phases.\cite{M.Naguib2011,M.Naguib2012} 
It has been repeatedly proven experimentally that 
by using the combination of acid treatment and sonication, the M$-$A bonds 
in some of the MAX phases can be broken while the M$-$X bonds are kept almost intact.
After the exfoliation process,
the A atoms are washed out from the MAX phase and the remaining set of single or multilayers is MXene. 
The crystal structures of derived 2D MXenes, M$_2$X, M$_3$X$_2$, M$_4$X$_3$, M$_5$X$_4$ 
M$_2$M$'$X$_2$, and M$_2$M$'_2$X$_3$, are shown schematically in Fig.~\ref{fig:maxenes}. 
So far, the single or multilayer 2D structures of 
Ti$_2$C,\cite{M.Naguib2012} 
Ti$_2$N,\cite{B.Soundiraraju2017}
V$_2$C,\cite{M.Naguib2013}
Nb$_2$C,\cite{M.Naguib2013}
Mo$_2$C,\cite{R.Meshkini2015}
Ti$_3$C$_2$,\cite{M.Naguib2011}
Zr$_3$C$_2$,\cite{J.Zhou2016} 
Nb$_4$C$_3$,\cite{M.Ghidiu2014} 
Ta$_4$C$_3$,\cite{M.Naguib2012} and 
Ti$_4$N$_3$\cite{P.Urbankowski2016} 
as well as 
TiNbC,\cite{M.Naguib2012}
(Ti$_{0.5}$Nb$_{0.5}$)$_2$C,\cite{M.Naguib2012}
(V$_{0.5}$Cr$_{0.5}$)$_3$C$_2$,\cite{M.Naguib2012}
Ti$_3$CN,\cite{M.Naguib2012} 
Mo$_2$TiC$_2$,\cite{B.Anasori2015_2}
Mo$_2$ScC$_2$,\cite{R.Meshkian2017}
Cr$_2$TiC$_2$,\cite{B.Anasori2015_2}
Mo$_2$Ti$_2$C$_3$,\cite{B.Anasori2015_2}
(Nb$_{0.8}$Ti$_{0.2}$)$_4$C$_3$,\cite{J.Yang2016_2} and  
(Nb$_{0.8}$Zr$_{0.2}$)$_4$C$_3$~\cite{J.Yang2016_2} 
have already been synthesized. 
The electronic properties of these 2D MXenes
are different from their corresponding MAX phases. 
It is also shown theoretically that upon a particular surface functionalization, some of the MXenes 
turn into semiconductors.\cite{M.Khazaei2013}
In contrast to the limited application of MAX phases as structural materials, 
many interesting electronic,\cite{M.Khazaei2013,M.Khazaei2014_3,H.Weng2015,M.Khazaei2016_2,M.Khazaei2016_1,M.Khazaei2015,Y.Liang2017} optical,\cite{H.Lashgari2014} mechanical,\cite{Z.Guo2015_1} 
magnetic,\cite{M.Khazaei2013,C.Si2015,J.He2016,M.Je2016,G.Gao2016,H.Kumar2017} and thermoelectric~\cite{M.Khazaei2013,M.Khazaei2014_3}, photocatalytic~\cite{Z.Guo2016} and many other applications~\cite{X.F.Yu2015_1,Y.Lee2015}
have been theoretically proposed for 2D MXenes. Experimentally, MXenes have attracted a lot of attention in the materials science community for
electronic transport,\cite{Y.Yang2017,S.Lai2015,F.Shahzad2016} energy storage,\cite{R.B.Rakhi2015,Y.Xie2014_1} energy conversion,\cite{H.Kim2017,H.Lin2016,G.Fan2017,J.Ran2017} 
and development of novel hybrid nanocomposites.\cite{X.Zhang2013,M.Xue2017} 
There are comprehensive reviews on the synthesis and application,\cite{B.Anasori2017_1} 
and the current theoretical status~\cite{M.Khazaei2017} of MXenes in the literature. 

It is now well accepted that the MAX phases are a great source for producing novel 2D systems. 
Considering the compositional varieties of MAX phases, a large number of 2D 
MXenes with unprecedented properties are expected to become available. 
Therefore, an important question to be answered at this moment is which of the 
MAX phases are promising candidates for a successful exfoliation into 2D materials. 
Theoretically, based on the analysis of tensile and
shear stresses, 
the mechanical exfoliation of various M$_2$AlC (M = Ti, Zr, Hf, V, Nb, Ta, Cr, Mo, and W) MAX phases into 2D M$_2$C MXenes have been investigated.\cite{Z.Guo2015} 
There, it was found that by applying a large
tensile strain, all the interlayer M$-$Al bonds can be broken, leading to the separation between 
M$_2$C and Al layers.\cite{Z.Guo2015} However, 
the chemical exfoliation process is a very complicated dynamical process, which is very difficult to model and simulate 
with all the details. 
Nevertheless, we can still gain great insights through studying the bond strength and the exfoliation energy because  
it is expected that MAX phases with weaker M$-$A bonds are promising for the successful exfoliation process. 
In this paper, by using a set of first-principles calculations based on density functional theory (DFT), we 
study the electronic structures, static exfoliation energies, force constants, 
and bond strengths of the 82 crystalline MAX phases. 
We show that the force constant is clearly correlated with the bond strength. 
We find that except for the MAX phases 
with the A elements of S or P, many of the MAX phases with other types of A elements have higher
possibility for exfoliation into 2D MXenes. Moreover, we examine the 
effect of charge injection on the bond strength of the MAX phases. 
For negatively charged MAX phases, the M$-$A bonds become elongated and weaker upon receiving electrons, 
thus facilitating the exfoliation process. 
However, the exfoliation becomes harder for positively charged MAX phases.

\section{METHODS OF CALCULATIONS}

Among different methods, first-principles calculations based on 
density functional theory (DFT) have proven to be reliable to predict various physical and chemical 
phenomena on the atomic scale. 
Hence, we employ DFT calculations to optimize the atomic structures and investigate the electronic structures of the MAX phases. 
All calculations are performed using the Vienna ab initio simulation package (VASP) code.\cite{vasp1996} 
The generalized gradient 
approximation (GGA) using the Perdew-Burke-Ernzerhof (PBE) 
functional is used to compute the exchange-correlation 
energy.\cite{pbe1996} 
The projected augmented wave approach with a plane wave cutoff 
energy of 520 eV is used to construct the wave functions. The atomic positions and lattice constants are 
fully optimized using the conjugate gradient method without imposing any symmetry. 
After structural optimization of atomic structures, the maximum 
residual force on each atom is less than 0.0001 eV/\AA. In the electronic self-consistency procedure, the total energies are converged to within
10$^{-8}$ eV/cell.  
For the optimization of M$_{n+1}$AX$_n$ (M$_2$M$'_{n}$AX$_{n+1}$), 
the Brillouin zone is sampled by taking 
18$\times$18$\times$9, 18$\times$18$\times$6, 18$\times$18$\times$3, and 18$\times$18$\times$1 
(18$\times$18$\times$6 and 18$\times$18$\times$3)
Monkhorst-Pack {\bf k} points~\cite{monkhorst1976} for $n=1$, 2, 3, and 4 (for $n=1$ and 2),
respectively. 
To calculate the projected density of states (PDOS), we use a larger number of  {\bf k} points, \textit{i.e.},  
24$\times$24$\times$12, 24$\times$24$\times$9, 24$\times$24$\times$6, and 24$\times$24$\times$3 
Monkhorst-Pack {\bf k} points for $n=1$, 2, 3, and 4, respectively. 
Spin-polarized calculations are employed to optimize the atomic structures of neutral and charged MAX phases. 
We find that most of the MAX phases are nonmagnetic, except for 
a few Cr-based MAX phases that exhibit weak magnetism. 
The total energies of all optimized MXenes are evaluated using spin-polarized calculations. 

The phonon calculations are carried out for nonmagnetic structures
using the PHONOPY package~\cite{phonopy2008} 
along with the VASP.\cite{vasp1996} 
The force constant is the second derivative of the total energy with respect to finite 
displacements of atoms $i$ and $j$ along the $x$, $y$, and $z$ directions, and is a 3$\times$3 matrix, 
given as one of the output results of the phonon calculations. 
Since the trace of the force constant matrix is independent of the coordinate system, \textit{i.e.}, 
invariant under a coordinate rotation,\cite{A.J.E.Foreman1957,Y.Liu2014}
following ref.\cite{Y.Liu2014} we consider the trace of the force constant matrix 
and refer to this scalar quantity as the force constant $F_{ij}$ between atoms $i$ and $j$ in the main text. 

Crystal orbital Hamilton population (COHP) analysis is a technique for partitioning 
the band-structure energy into bonding, nonbonding, 
and antibonding contributions of the localized atomic basis sets.\cite{R.Dronskowski1993} 
Alternatively, the COHP may be described as a hopping-weighted density-of-states between a pair of adjacent 
atoms defined as 
\begin{equation}
\small
{\rm COHP}_{\mu\vec{T},\nu\vec{T'}}(E)=H_{\mu\vec{T},\nu\vec{T'}}\sum_{j,\vec{\bf k}}f_j(\vec{\bf k})C^\ast_{\mu\vec{T},j}  (\vec{\bf k}) C_{\nu\vec{T'},j} (\vec{\bf k}) \delta(\epsilon_j(\vec{\bf k})-E),
\label{eq:COHP}
\end{equation} 
where $H_{\mu\vec{T},\nu\vec{T'}}$ is the matrix element of the Hamiltonian matrix $H$ and 
$C_{\mu\vec{T},j}(\vec{\bf k})$ represents the eigenvector coefficient of an atomic orbital $\mu$. 
$\epsilon_j(\vec{\bf k})$ is the $j$th band energy at momentum $\vec{\bf k}$, 
$\vec{T}$ is the translational lattice vector, and $f_j(\vec{\bf k})$ is the occupation number of that state. 
Similar to the density of states (DOS) where the energy integration up to the Fermi energy gives the number 
of electrons, the energy integration of the COHP up to a particular band energy for a pair of atoms indicates 
their bond strength in terms of their contribution to this band structure energy. 
Upon integrating {\em all} COHPs including the on-site elements that correspond to a self-interaction of an atom, 
up to the Fermi energy, 
one is left with the integral of the density-of-energy (DOE) function, 
another partitioning of the band-structure energy.\cite{M.Kupers2017} 
All the COHP calculations are done using the Local Orbital Basis Suite Towards Electronic-Structure 
Reconstruction (LOBSTER) code~\cite{V.L.Deringer2011,S.Maintz2013,S.Maintz2016} with the 
pbeVaspFit2015 basis set.\cite{S.Maintz2016}

The bond order ($N$) is calculated based on the Pauling classical descriptor 
$D_N = D_1 - \gamma \AA \times \log_{10} N$,
where $D_1$ and $D_N$ are the length of a single bond and the measured 
bond length, respectively.\cite{L.Pauling1947,L.Pauling1986} $D_1$ for a X$-$Y bond is given 
as $D_1 = R_X+R_Y$, where $R_X$ and $R_Y$  are the single-bond radii of elements X and Y. 
In this formula, the coefficient $\gamma$
can vary in the range between 0.6 and 0.7.\cite{L.Pauling1986} 
In our calculations, we set that $\gamma=0.6$ in consistency with the previous studies in the literature.\cite{V.L.Deringer2015,N.W.Tideswell1957} 
The lengths of single-bond radii of pure metals have already been tabulated for most 
elements.\cite{L.Pauling1947,L.Pauling1986,P.Pyykko2009}

The elastic constant C$_{33}$  is obtained by parabolically fitting the total energy of the strained crystal 
along the $c$ direction with respect to small distortions 
($\Delta c/c =\pm$0.02, $\pm0.015$, $\pm0.01$, and $\pm0.005$).\cite{L.Fast1995}
The static exfoliation energy $E_{\rm Exfoliation}$ of a bulk MAX phase into 2D MXenes is calculated 
through E$_{\rm Exfoliation}$= -[E$_{\rm tot}({\rm MAX\, phase}$)- 2 E$_{\rm tot}({\rm MXene}$)- 2 E$_{\rm tot}({\rm A}$)]/$(4S)$, 
where $E_{\rm tot}({\rm MAX\, phase})$, $E_{\rm tot}({\rm MXene})$ and $E_{\rm tot}({\rm A})$ stand for 
the total energies of bulk MAX phase, 2D MXene, and A element, respectively,\cite{M.Khazaei2014_2}. 
Here, $S=\sqrt{3}a^2/2$ is the surface area and $a$ is the lattice parameter of the MAX phase. 
Exfoliating a MAX phase, 
each unit cell of the MAX phase generates two MXene layers with totally 4 surfaces. 
Hence, the exfoliation energy is divided by 4 in the above formula.
The total energy of A element, $E_{\rm tot}({\rm A})$, is 
estimated from its most stable bulk structure.  

\begin{figure}[t]
\centering
  \includegraphics[scale=0.36]{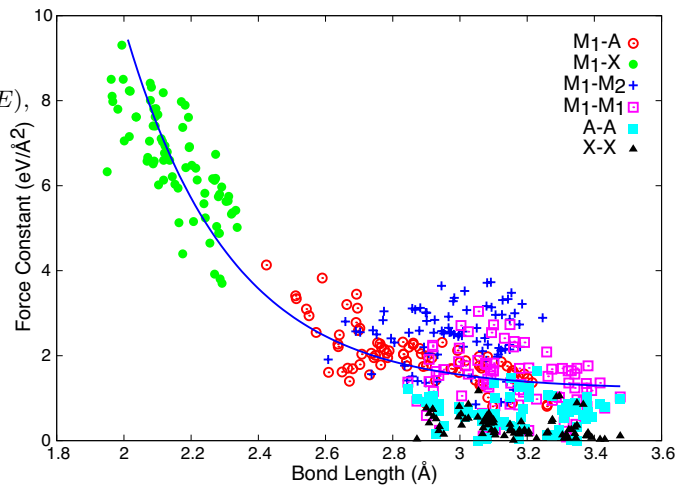}
  \caption{Calculated force constants of different bonds versus the corresponding bond lengths 
  for the 82 experimentally synthesized MAX phases. The line is a guide to the eye.}
  \label{fig:bonddistance}
\end{figure}

\section{RESULTS AND DISCUSSION}

In experiments, the exfoliation process occurs dynamically and there are many parameters to control the process, 
such as the acidic solution type, the concentration of the acid, and the temperature. However, it is 
very difficult, if not impossible, to theoretically consider all these details of the process using the 
currently available computational resources, although there have been several attempts to look into 
such a complicated process in the literature.\cite{P.Srivastava2016,A.Mishra2016}
Therefore, currently, the static calculations are considered as the only straightforward way to provide useful 
information on the exfoliation possibility. In this regard, force constants, integrated crystal orbital Hamilton 
population (ICOHP), bond orders, and 
exfoliation energies are important quantities that can provide quantitative measures of the bond strength 
and the ease of exfoliation. This simulation-based approach is highly valuable since  
the bond strength is not an experimentally measurable quantity 
(except for diatomic molecules) but nonetheless it is chemically meaningful. 

\begin{table}
\caption{List of MAX phases exfoliated experimentally into MXenes and the calculated  
exfoliation energy E$_{\rm Exfoliation}$ (eV/\AA$^2$) and total force constant FC$_i$ (eV/\AA$^2$) for an 
$i$ (= M, A, or X) atom 
counting all contributions from the neighboring atoms. 
}
\begin{tabular}{lcccc}
\hline
MAX Phase                                                  &    FC$_{\rm M}$              & FC$_{\rm A}$             & FC$_{\rm X}$            & E$_{\rm Exfoliation}$  \\
\hline
Ti$_2$AlC\cite{M.Naguib2012}                       & 46.742                    &      15.078           &     51.518        &       0.164        \\
Ti$_2$AlN\cite{B.Soundiraraju2017}                       &    45.139                           &      18.813                       &  50.626                       &               0.190           \\
V$_2$AlC\cite{M.Naguib2013}                       &  51.943                   &      21.855          &      59.168       &      0.205       \\
Nb$_2$AlC\cite{M.Naguib2013}                     &       51.706                     &       18.205              &      49.187                &        0.185        \\
Ti$_3$AlC$_2$\cite{M.Naguib2011}              &   49.175                  &      14.708           &     49.514        &       0.158        \\
 Zr$_3$AlC$_2$\cite{J.Zhou2016}              &    46.019                 &       11.886          &     40.511        &        0.131       \\
 Ti$_4$AlN$_3$\cite{P.Urbankowski2016}               &     48.502                &       19.080          &     51.258        &        0.193       \\
Nb$_4$AlC$_3$\cite{M.Ghidiu2014}             &      53.323               &      16.680           &     48.319        &        0.175       \\
Ta$_4$AlC$_3$\cite{M.Naguib2012}              &         53.959               &       18.618                      &       55.066               &            0.197            \\
Mo$_2$ScAlC$_2$\cite{R.Meshkian2017}         &       45.170              &         20.251         &     44.822       &        0.201       \\
 Mo$_2$TiAlC$_2$\cite{B.Anasori2015_2}         &       49.797              &       20.086           &     52.324        &       0.196        \\
 Cr$_2$TiAlC$_2$\cite{B.Anasori2015_2}          &      44.253               &        20.124         &      53.691       &        0.184       \\
Mo$_2$Ti$_2$AlC$_3$\cite{B.Anasori2015_2}  &      51.522               &        20.094         &     52.379        &      0.199          \\
\hline
\end{tabular}
\label{tab:experiment}
\end{table}

\subsection{Force constant analysis}

The force constant is the second derivative of the total energy with respect to the displacement of an atom, 
which is also equivalent to the first derivative of the force. 
It represents the force required to displace an atom by an infinitesimal amount 
while all other atoms are held fixed in their equilibrium positions. 
Because of this, the force constant 
has been used as a bond strength descriptor 
in the literature.\cite{K.Brandhorst2008,D.Cremer2010,R.G.Pearson1993}
Historically, the force constant has been used to describe the strength of a bond 
in terms of dissociation, particularly applied to investigate the dissociation energy of atomic 
dimers.\cite{J.R.Lombardi2002}
While the binding energy is the energy required to dissociate a bond from its equilibrium position, 
the force constant is the curvature of the energy vs. bond length curve at the equilibrium position. 
Although in general the binding energy and the force constant are not mathematically related directly, 
it is easy to conceive that when the binding is weak, the curvature of the energy vs. bond length curve 
is small, and vice versa, namely, when the curvature is large, 
the binding is strong. 
A very simple justification for this relation between the binding energy and the 
force constant can be found when the energy scale  of the interatomic interaction potential has 
a Morse~\cite{burgi1987} or Lennard-Jones potential form.\cite{H.S.Johnston1964} 
While both quantities can easily be calculated for molecules, in the case of solids, it is impossible to consider 
the energy change when only one bond is stretched, but we rather consider the cohesive energy 
where {\it all} bonds are stretched. 
Therefore, it is preferable to discuss the bond strength based on the force constant, 
which one can calculate theoretically even for solids. 
Moreover, in recent years, the concept of a force constant descriptor 
has been developed to investigate the strength of interactions in solids~\cite{Y.Liu2014,J.Hong2016} and 
interfaces.\cite{O.V.Pupysheva2010,L.Wang2017}
For the case of Sb$_2$Se$_3$, for example, it was found that there is 
a clear correlation between the stiffness of a particular bond (expressed by the force constant) and 
the covalency (expressed by COHP analysis).\cite{V.L.Deringer2015} 
This motivates us to investigate the bond strength in
the MAX phases from the view point of the force constant. 

\begin{table}
\caption{List of MAX phases sorted with FC$_{\rm A}$ ($\leq$ 21.855 eV/\AA$^2$) 
in ascending order (from top left to bottom right). 
The calculated FC$_{\rm A}$ (eV/\AA$^2$) is shown in each parenthesis.
The results for all studied MAX phases are provided in the supporting information file.}
\small
\begin{tabular}{ccc}
\hline
Sc$_2$GaC (5.590) &
Sc$_2$AlC  (6.412)&
Sc$_2$InC	(8.771)\\
Sc$_2$TlC	(9.628)&
Ti$_2$CdC	(11.683)&
Zr$_2$AlC	(11.852)\\
Zr$_3$AlC$_2$	(11.886)&
Ti$_3$AuC$_2$	(12.108)&
Ti$_5$AlC$_4$	(12.395)\\
Zr$_2$InC	(13.475)&
Zr$_2$TlC	(13.693)&
Hf$_2$AlC	(13.999)\\
Ti$_2$GaC	(14.594)&
Ti$_4$GaC$_3$	(14.668)&
Ti$_3$AlC$_2$	(14.708)\\
Zr$_2$AlN	(15.047)&
Ti$_2$AlC	(15.077)&
Hf$_2$InC	(15.64)\\
Nb$_5$AlC$_4$	(15.724)&
Hf$_2$TlC	(16.117)&
Zr$_2$InN	(16.327)\\
Zr$_2$TlN	(16.380)&
Zr$_2$PbC	(16.658)&
Nb$_4$AlC$_3$	(16.680)\\
Ti$_2$InC	(17.174)&
Hf$_2$AlN	(17.288)&
Ti$_2$TlC	(17.438)\\
Nb$_2$GaC	(17.520)&
Zr$_2$SnC	(17.769)&
Ti$_2$GaN	(17.942)\\
Nb$_2$AlC	(18.205)&
Zr$_2$BiC	(18.344)&
Ta$_5$AlC$_4$ (18.364)\\
Hf$_2$PbC	(18.518)&
Ta$_4$AlC$_3$	(18.618)&
Ti$_3$SiC$_2$	(18.786)\\
Ti$_4$SiC$_3$	(18.801)&
Ti$_2$AlN	 (18.813)&
Ti$_3$GeC$_2$	(18.867)\\
Nb$_2$InC	(18.884)&
Ti$_2$GeC	(19.016)&
Ti$_4$AlN$_3$	(19.08)\\
Ti$_4$GeC$_3$	(19.118)&
Hf$_2$SnC	(19.217)&
Mo$_2$GaC	(19.247)\\
Ti$_2$SiC	(19.366)&
Ta$_2$GaC	(19.464)&
Cr$_2$GaN	(19.492)\\
Ti$_3$IrC$_2$	(19.702)&
V$_2$GaC	(19.803)&
Ti$_2$InN	(19.971)\\
Mo$_2$TiAlC$_2$	(20.086)&
Mo$_2$Ti$_2$AlC$_3$	(20.094)&
V$_2$GaN	(20.107)\\
Cr$_2$TiAlC$_2$ (20.124)&
Mo$_2$ScAlC$_2$ (20.251)&
Ta$_2$AlC$_2$	(20.536)\\
Ti$_2$PbC	(20.543)&
Ta$_2$AlC	(20.596)&
Cr$_2$GaC	(20.775)\\
V$_4$AlC$_3$	(20.866)&
V$_3$AlC$_2$	(20.894)&
Hf$_2$SnN	(21.213)\\
Ti$_2$SnC	(21.586)&
V$_2$AlC	(21.855)\\
\hline
\end{tabular}
\label{tab:FCA}
\end{table}

In order to evaluate the force constants of the bonds in MAX phases, we perform a set of phonon calculations 
on the 82 experimentally crystalline MAX phases, listed in Table 1S. 
As expected and consistent with the experimental results,
no negative phonon frequencies are found in the phonon spectra of these MAX phases. 
This indicates that these MAX phases are 
dynamically stable. The phonon spectra calculated here are given in the supporting information file. 
Table 1S includes the results of the force constants for 
M$_1-$M$_1$, M$_1-$M$_2$, M$_1-$X, M$_1-$A, A$-$A, and X$-$X bonds. 
 Here, M$_1$ and M$_2$ represent transition metals belonging to the first and second 
layers of transition metals close to the A element, 
and the X layer is sandwiched between M$_1$ and M$_2$ layers  (see Fig.~\ref{fig:maxphases}).
In M$_2$AX, M$_3$AX$_2$, M$_4$AX$_3$, and M$_5$AX$_4$ MAX phases,
the transition metals of M$_1$ and M$_2$ are of the same type, while in M$_2$M$'$AX$_2$ and 
M$_2$M$'_2$AX$_3$ MAX phases, M$_1$ and M$_2$ are of different types. 
The results of the force constants for these bonds are summarized in Fig.~\ref{fig:bonddistance}. 
A general trend can be revealed that 
shorter bonds are stiffer. It is also observed that 
the force constants of the M$_1-$X bonds are significantly larger than those of other bonds. 
This indicates that the M$_1-$X bonds are the strongest in the 
MAX phases, which is the main reason for the stability of these MAX phases and the resulting MXenes. 
Since the force constants of the M$_1-$A bonds are smaller than those of the M$_1-$X bonds, implying that 
the M$_1-$A bonds are weaker than the M$_1-$X bonds, 
it is expected that the bulk modulus of MAX phases is not larger than that of their corresponding
binary MX compounds. 
This is consistent with previous studies where the bulk modulus of 
M$_2$AX MAX phases was classified into two groups, a group with a similar bulk modulus and a group with 
a smaller bulk modulus as compared to that of the binary MX compounds.\cite{Z.Sun2004,D.Music2006}

\begin{table}
\caption{List of MAX phases sorted by FC$_{\rm X}$ ($\geq$40.511 eV/\AA$^2$) 
in descending order (from top left to bottom right). 
The calculated FC$_{\rm X}$ (eV/\AA$^2$) is shown in each parenthesis.
The results for all studied MAX phases are provided in the supporting information file.}
\small
\begin{tabular}{ccc}
\hline
V$_2$AlC	(59.168)&
V$_2$GaC (58.198)&
V$_4$AlC$_3$	(58.123)\\
Cr$_2$AlC (57.783)&
Ta$_2$AlC (57.183)&
Ta$_3$AlC$_2$ (55.922)\\
V$_3$AlC$_2$	(55.760)&
Ta$_2$GaC (55.684)&
Ta$_4$AlC$_3$ (55.066)\\
V$_2$SiC	 (54.431)&
Cr$_2$GaC (53.940)&
Cr$_2$TiAlC$_2$ (53.691)\\
Cr$_2$GeC (53.649)&
Mo$_2$Ti$_2$AlC$_3$ (52.379)&
Mo$_2$TiAlC$_2$ (52.324)\\
Ta$_5$AlC$_4$(53.042)&
V$_2$GeC (52.022)&
Ti$_2$AlC	(51.518)\\
Ti$_2$SiC	(51.315)&
V$_3$SiC$_2$	(51.296)&
Ti$_4$AlN$_3$	(51.258)\\
Ti$_2$GaC (50.935)&
Ti$_3$IrC$_2$ (50.703)&
Ti$_2$AlN (50.626)\\
Ti$_2$GaN (50.307)&
V$_2$PC (49.711)&
Ti$_3$SiC$_2$	(49.697)\\
Ti$_3$AlC$_2$	(49.514)&
Ti$_2$GeC (49.400)&
Nb$_2$AlC (49.187)\\
Ti$_2$GeC$_2$ (48.641)&
Nb$_2$GaC (48.560)&
Nb$_4$AlC$_3$ (48.319)\\
V$_2$GaN (48.010)&
Ti$_4$SiC$_3$	(47.710)&
Ti$_3$AuC$_2$ (47.647)\\
Ti$_2$CdC (47.573)&
Ti$_4$GaC$_3$ (47.563)&
Ti$_2$InC	(47.441)\\
V$_2$AsC (47.374)&
Cr$_2$GaN (47.339)&
Hf$_2$AlC (47.038)\\
Ti$_4$GeC$_3$ (46.894)&
Mo$_2$GaC (46.103)&
Ti$_2$TlC	(45.923)\\
Ti$_5$AlC$_4$	(45.731)&
Ti$_3$SnC$_2$ (45.549)&
Hf$_2$InC (45.447)\\
Nb$_5$AlC$_4$ (45.336)&
Nb$_2$InC (45.318)&
Ti$_2$SC (45.014)\\
Mo$_2$ScAlC$_2$(44.822)&
Hf$_2$TlC (44.767)&
Ti$_2$SnC (44.679)\\
Ti$_2$InN	(44.642)&
Hf$_2$SC	(44.560)&
Hf$_2$SnC (43.319)\\
Nb$_2$SC (42.841)&
Hf$_2$AlN (42.212)&
Hf$_2$PbC (41.947)\\
Ti$_2$PbC (41.906)&
Nb$_2$PC (41.902)&
Nb$_2$SnC (41.843)\\
Zr$_2$AlC (41.725)&
Nb$_3$SiC$_2$ (41.391)&
Ti$_2$AsC (41.355)\\
Nb$_2$AsC (40.670)&
Zr$_2$InC (40.623)&
Zr$_3$AlC$_2$ (40.511)\\
\hline
\end{tabular}
\label{tab:FCX}
\end{table}

As shown in Fig.~\ref{fig:bonddistance}, the force constants of the M$_1-$M$_2$ bonds are as large as 
those of the M$_1-$A bonds but smaller than those for the M$_1-$X bonds. 
This suggests that the M$_1-$M$_2$ bonds are also weaker or softer than the M$_1-$X bonds, 
which is in agreement with chemical intuition. 
However, the M$_1-$A and M$_1-$M$_2$ bonds are still strong enough to contribute to the stability 
of MAX phases. 
Because of the periodicity of the structure, the lengths of the M$_1-$M$_1$, A$-$A, 
and X$-$X bonds are equal to the lattice constant \textit{a}, which is larger than the lengths of the 
M$_1-$X, M$_1-$A, or M$_1-$M$_2$ bonds. 
Since the X$-$X interatomic distances are more than two times larger than the atomic radii of X elements, 
the orbital overlap between the X atoms in the X$-$X bonds should be weak. 
On the other hand, twice the atomic radii of A or M atoms are comparable with the A$-$A or M$_1-$M$_1$ 
bond lengths, suggesting that the A or M atoms in  the A$-$A or M$_1-$M$_1$ bonds exhibit strong orbital overlap. 
Therefore, we expect the X$-$X bonds to be the weakest among all the bonds, as observed 
in Fig.~\ref{fig:bonddistance} and Table 1S.

In order to exfoliate MAX phases into 2D MXenes, the bonds between an A atom and its neighboring A or M 
atoms have to be broken. In other words, to detach an A atom from the bulk MAX phase, 
at least six A$-$A bonds and six M$_1-$A bonds must be broken. 
Hence, to propose which of the MAX phases can be exfoliated to MXenes, here we calculate the total 
force constants for an $i$ (= M, A, or X) atom counting all contributions from other atoms, \textit{i.e.}, 
FC$_{i}$=$\sum_j$FC$_{ij}$, where FC$_{ij}$ is the force constant between atoms $i$ and $j$, and 
$j$ enumerates all atoms in the supercell of the phonon calculation except atom $i$. 
Generally, the force constant is a $3\times3$ second-order tensor and here we consider the trace of this tensor 
as FC$_{ij}$.
Therefore, FC$_{i}$ is related to the second derivative of the total energy $E$ with respect to displacement 
$r_i^{(\alpha)}$ (where $\alpha=x,y, z$) of atom $i$ in the $\alpha$ direction, \textit{i.e.}, 
FC$_{i}$=$\sum_{\alpha}\partial^2E$/$\partial^2r_i^{(\alpha)}$. 
After the systematic calculation (see the results in Table 1S), 
we find that in most of the MAX phases the force constants between 
an A atom and the surrounding atoms located further than the first shell of neighbors are weak. 
Therefore, an A atom mainly interacts with the neighboring atoms in its first shell of neighbors, which 
are of the M or A atoms. 
We also find that the total force constants of an X or M atom are significantly larger than those of an A atom 
(see Table~\ref{tab:experiment} and Table 1S). This explains why in the exfoliation process the bonds formed by
 M and A atoms are typically broken rather than those formed by M and X atoms.    

\begin{figure}[t]
\centering
  \includegraphics[scale=0.35]{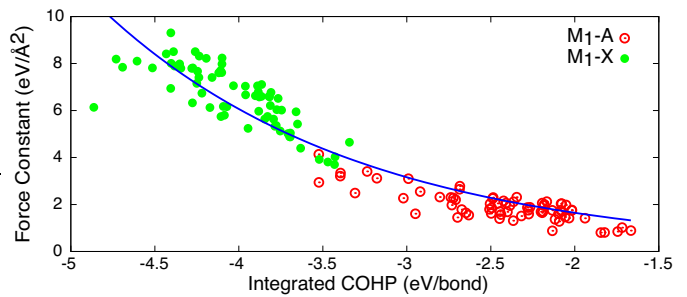}
  \caption{
  Bond strength versus force constant for the M$_1-$A and M$_1-$X bonds for various experimentally 
  synthesized MAX phases.
  Here the bond strength is quantified by the integrated crystal orbital Hamilton population (COHP) up to the 
  Fermi energy over all the atomic orbital interactions between the atoms forming the bonds.
  The line is a guide to the eye.}
  \label{fig:bondhybrid}
\end{figure}

In order to examine which of the MAX phases are well-suited for the successful exfoliation into 
2D MXenes, the total force constant FC$_{\rm A}$ for the A elements is a very useful quantity. 
As a simple criterion, we can expect that MAX phases with smaller FC$_{\rm A}$ can be considered 
as promising candidates for the exfoliation process. 
In this respect, Ti$_2$AlC, V$_2$AlC, Ti$_3$AlC$_2$, Zr$_3$AlC$_2$, Ti$_4$AlN$_3$, Nb$_4$AlC$_3$, Mo$_2$TiAlC$_2$, Cr$_2$TiAlC$_2$, Mo$_2$ScAlC$_2$, and 
Mo$_2$Ti$_2$AlC$_3$ are the list of MAX phases that have already been exfoliated to Ti$_2$C, V$_2$C, Ti$_3$C$_2$, Zr$_3$C$_2$, Ti$_4$N$_3$, Nb$_4$C$_3$, Mo$_2$TiC$_2$, 
Cr$_2$TiC$_2$, Mo$_2$ScC$_2$, and Mo$_2$Ti$_2$C$_3$, respectively. 
All these compounds contain Al in the MAX phases. 
The calculated FC$_{\rm A}$ values in these compounds are summarized in Table~\ref{tab:experiment}. 
As shown in Table~\ref{tab:experiment}, 
the minimum (maximum) FC$_{\rm A}$ is for Zr$_3$AlC$_2$ (V$_2$AlC). 
Therefore, we expect that by using appropriate acids in the experiments it might be possible to break 
the bonds as strong as the V$-$Al bond. 
Based on this criterion, we screen our data in Table 1S for compounds 
with FC$_{\rm A}$ less than 21.855 eV/\AA$^2$ and the results are summarized in Table~\ref{tab:FCA}. 
The compounds listed in Table~\ref{tab:FCA} are sorted based on their FC$_{\rm A}$ 
values in ascending order: the smaller FC$_{\rm A}$ is, the higher the chance for successful exfoliation becomes. 
Therefore, the table lists the compounds from the most promising candidate to the least promising one for
successful exfoliation. However, it is noted that all compounds listed in Table~\ref{tab:FCA} have better potential 
to be exfoliated into MXenes because 
their respective FC$_{\rm A}$ is below the upper limit of 21.855 eV/\AA$^2$.  

The exfoliation process would be successful experimentally only if the obtained 2D MXenes are 
structurally perfect. This requires that the M$_1-$M$_1$, M$_1-$M$_2$, and M$_1-$X bonds should be 
stronger than the M$_1-$A and A$-$A bonds. 
Otherwise, during the exfoliation process, other bonds may also be broken besides the M$_1-$A bonds. 
In such a case, either the MXenes would not be formed at all or they would be formed with many M and X defects. 
Since the M$_1-$X bonds are the strongest in MAX phases, we calculate the total force constant FC$_{\rm X}$ 
for the X elements in Table.~\ref{tab:experiment} for the MAX phases experimentally exfoliated into MXenes. 
We find that the maximum (minimum) of FC$_{\rm X}$ is for V$_2$AlC (Zr$_3$AlC$_2$). 
Therefore, in order to successfully synthesize MXenes, 
FC$_{\rm X}$ should at least be as large as that for Zr$_3$AlC$_2$ (40.511 eV/\AA$^2$). 
Based on this criterion, we re-screen the results of FC$_{\rm X}$ in all MAX phases listed in Table 2S
and the results are summarized in Table~\ref{tab:FCX}. 
The compounds listed in Table~\ref{tab:FCX} are sorted based on their FC$_{\rm X}$ values 
in descending order: the larger FC$_{\rm X}$ is, the higher the quality of the obtained MXene is expected 
once it is successfully exfoliated. 

\begin{figure}[t]
\centering
  \includegraphics[scale=0.35]{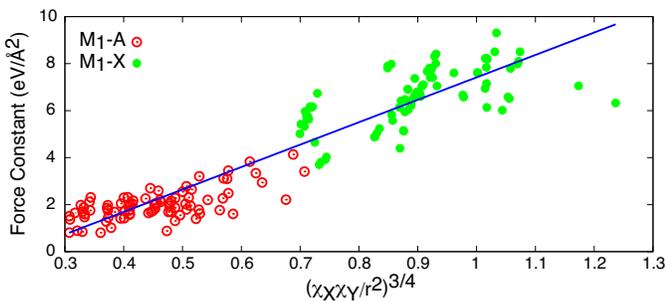}
  \caption{
  Force constant of the M$_1-$A and M$_1-$X bonds as a function of electronegativities 
  for various experimentally synthesized MAX phases. 
  $\chi_X$ and $\chi_Y$, and $r$ are the electronegativities of $X$ and $Y$ atoms, and the length 
  of the $X-Y$ bond. 
  The straight line is a guide to the eye.}
  \label{fig:bondionic}
\end{figure}

The results shown in Table 1S are also helpful to predict which of the MAX phases have the least chance 
for successful exfoliation into 2D MXenes.  
It is found in Table 1S that in MAX phases with the A elements being P or S, 
the M$_1-$P or M$_1-$S bonds are as strong as the M$_1-$X bonds. 
This suggests that charge neutral MAX phases with A = S or P are typically less promising compounds for 
the exfoliation process. 
Another important implication of the results shown in Table~\ref{tab:FCA} is a possibility of MAX phases 
for ion intercalation and their use for applications such as Li ion batteries.\cite{J.Xu2017,A.T.Tesfaye2017} 
Because of the weaker bond strength of the M$_1-$A bonds as compared with the M$_1-$X bonds,
and the greater interatomic distances, MAX phases possess large free space near A atoms.
It can be inferred that the guest ions should be intercalated near the A atoms. 
Therefore, MAX phases with weaker strength of the M$_1-$A bonds might find 
better applications in Li ion batteries.

\subsection{Bond strength analysis}

From a chemical point of view, the bond strength of a covalent interaction depends on the overlap between 
the atomic orbitals of the two atoms that form a bond. Somewhat simplified, the polarity of the bond 
results from the electronegativity difference between the two atoms connected by the bond~\cite{D.Cremer2000}. 
Typically, the overlap decreases exponentially with the bond length.\cite{D.Cremer2000} 
The electronegativity difference of the two atoms forming the bond also 
determines the effective charges on the atoms in the $\frac{1}{r}$ potential.\cite{D.Cremer2000} 
As explained in the section of methods of calculations, the integrated crystal orbital Hamilton population 
(COHP) up to the Fermi energy 
measures the mixing (\textit{i.e.}, the amount of interactions) of atomic orbitals, 
eventually determining the bond strength, and the COHP method has been widely 
applied to a plethora of 
materials,\cite{dronskowski2005} including MAX phases.\cite{D.Music2007}

Here, we calculate the integrated COHP (ICOHP) up to the Fermi energy over all the atomic orbital interactions
between the atoms forming the bonds. 
Figure~\ref{fig:bondhybrid} shows the force constant versus the bond strength 
quantified by the ICOHP for the M$_1-$X and M$_1-$A bonds. 
Note that the more negative (\textit{i.e.}, the more energy-lowering)  the ICOHP 
is, the stronger the bond is. 
It is clearly observed in Fig.~\ref{fig:bondhybrid} that the force constant intimately correlates with the strength of 
bonds: the force constant increases with increasing the bond strength. 
It is also found that the mixing 
of orbitals in the M$_1-$X bonds is larger than that in the M$_1-$A bonds.  
Both M$_1-$X and M$_1-$A bonds are polar because the electronegativities $\chi_{M_1}$, $\chi_A$, and $\chi_X$
of the M$_1$, A, and X elements are 
different. Generally, the polarity of the M$_1-$X bonds (\textit{i.e.}, $\chi_{M_1}-\chi_X$) is larger than that of the 
M$_1-$A bonds (\textit{i.e.}, $\chi_{M_1}-\chi_A$),  
Therefore, in addition to the orbital interaction, 
the ionic effect may also play a role in determining the difference of the stiffness between the 
M$_1-$X and M$_1-$A bonds. 

 \begin{figure}[t]
\centering
  \includegraphics[scale=0.35]{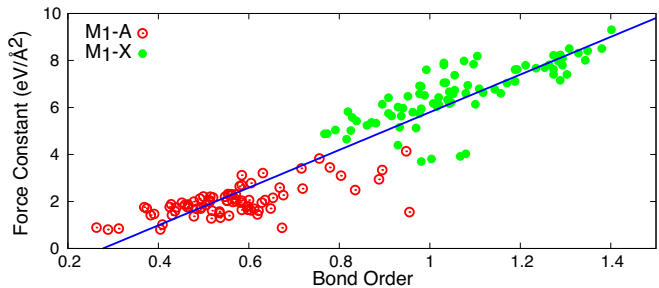}
  \caption{Force constant of the M$_1-$A and M$_1-$X bonds as a function of the corresponding 
  Pauling bond order, calculated for various experimentally synthesized MAX phases. 
  The straight line is a guide to the eye.}
  \label{fig:bondorder}
\end{figure}

There is no well-developed formalism to isolate the contribution of polarity of the bond to the force constant,  
except for extracting the ionicity (as well as the orbital mixing) from modern DFT-type PAW wave 
functions,  \textit{e.g.}, in the search for advanced materials via materials mapping.\cite{M.Esser2017}
Here, we rely on an empirical model proposed for molecules,\cite{ W.Gordy1946,W.Gordy1956,S.Kaya2016} 
 \textit{i.e.}, the force constant ($k$) of diatomic molecules, $k=aN(\chi_X \chi_Y/r^2)^{3/4}+b$, 
where $a$ and $b$ are constants, $\chi_X$ and $\chi_Y$ stand for the electronegativities of X and Y atoms, 
respectively, $r$ is the X$-$Y bond length, and
$N$ is the number of bonds (bond order, $N=1$, 2, and 3) between 
$X$ and $Y$ atoms.\cite{W.Gordy1946,W.Gordy1956} 
The physics behind the empirical model can be easily understood because the 
amount of charge on the nuclei is related to the electronegativities of the the bonded atoms, 
and the force $F$ between a nucleus and an electron in the bonded atoms 
can be estimated from the simple rule of electrostatic point charge.\cite{A.L.Allred1958} 
From Hooke's law $F=-k\Delta r$, the attractive force between the two bonded atoms is related 
to the force constant $k$, where $\Delta r$ is the change in the bond length.  
Hence, the overall dependence of the force constant on the electronegativities 
and the bond distance can be justified. 
In this empirical model, the parameters $a$, $N$, and $b$ are found through a fitting process. 
Figure~\ref{fig:bondionic} shows the force constant, calculated in Fig.~\ref{fig:bonddistance}, as a function of 
$(\chi_X\chi_Y/r^2)^{3/4}$.
Although the empirical model is a very rough approximation, it can still capture the physics of the polar bond in 
MAX phases. We find that $(\chi_X\chi_Y/r^2)^{3/4}$ varies almost linearly 
with the force constant and it is larger for the M$_1-$X bonds than the M$_1-$A bonds. 
The latter indicates that in addition to the orbital interaction, the polarity plays 
an important role in the stiffness of the M$_1-$X bonds.

\subsection{Bond order analysis}

In chemistry, the total number of bonds ($N=1$, $2$, and $3$) between bonded atoms, 
the so-called bond order, is usually assigned to represent the strength of a bond. 
It is considered that the bond becomes stronger with increasing bond order. 
For instance, in the literature, the carbon-carbon bonds are categorized as single, double, or triple-bonds 
in various compounds when the carbon-carbon bond distances are around 1.542, 1.339, and 1.204 \AA, 
respectively.
As described in the methodology section, Pauling has proposed an empirical 
logarithmic model to describe the bond order.\cite{L.Pauling1947,L.Pauling1986} 
Based on the Pauling formula along with the metallic radii of elements provided in 
Ref.~\cite{L.Pauling1986}, we calculate in Fig.~\ref{fig:bondorder} the force constants of 
the M$_1-$A and M$_1-$X bonds with respect to their corresponding bond orders 
in various MAX phases. 
It is clearly found that as the bond order increases, the force constant also increases. 
The bond orders for the M$_1-$X bonds are larger than that for the M$_1-$A bonds, 
thus indicating that the M$_1-$X bonds are stronger than the M$_1-$A bonds, which is consistent with 
the COHP analysis given above. 
Moreover, it is observed in Fig.~\ref{fig:bondorder} that the bond orders for the M$_1-$A bonds 
are less than 1. This suggests that the bond strength of the M$_1-$A bonds is weaker than the single bonds 
in pure metals of A and M. We have repeated the same analysis using the single bond radii given in 
Ref.~\cite{P.Pyykko2009} and obtained the same conclusion (see supplementary data).   

  \begin{figure}[t]
\centering
  \includegraphics[scale=0.35]{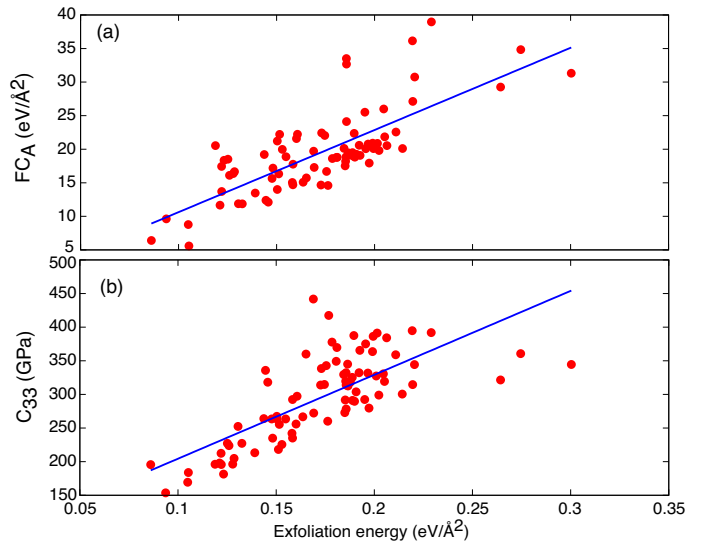}
  \caption{ (a) Force constant  FC$_{\rm A}$ for the A atoms and (b) elastic constant C$_{33}$ 
  along the $c$ direction 
  versus exfoliation energy $E_{\rm Exfoliation}$ (eV/\AA$^2$), calculated for various experimentally 
  synthesized MAX phases.
  The straight lines in (a) and (b) are guides to the eye.
  }
  \label{fig:exfoliation}
\end{figure}

\begin{table}
\caption{
List of MAX phases sorted by the exfoliation energy $E_{\rm Exfoliation}$ (eV/\AA$^2$) 
($\le 0.205$ eV/\AA$^2$) in ascending order (from top left to bottom right). 
The results for all studied MAX phases are provided in the supporting information file.
}
\small
\begin{tabular}{ccc}
\hline
Sc$_2$AlC	(0.086)&
Sc$_2$TlC	(0.094)&
Sc$_2$InC	(0.105)\\
Sc$_2$GaC	(0.105)&
Ti$_2$PbC	(0.119)&
Ti$_2$CdC	(0.121)\\
Ti$_2$TlC	        (0.122)&
Zr$_2$TlC	(0.122)&
Zr$_2$BiC	(0.123)\\
Hf$_2$PbC	(0.125)&
Hf$_2$TlC	(0.126)&
Zr$_2$TlN	(0.128)\\
Zr$_2$PbC	(0.129)&
Zr$_3$AlC$_2$	(0.131)&
Zr$_2$AlC	(0.133)\\
Zr$_2$InC	(0.139)&
Hf$_2$SnC	(0.144)&
Ti$_5$AlC$_4$	(0.145)\\
Ti$_3$AuC$_2$	(0.146)&
Hf$_2$InC	(0.148)&
Ti$_2$InC	 (0.148)\\
Hf$_2$AlC	(0.150)&
Hf$_2$SnN	(0.150)&
Zr$_2$InN	(0.151)\\
Nb$_2$SnC	(0.152)&
Ti$_2$InN	        (0.153)&
Nb$_2$InC	(0.155)\\
Zr$_2$AlN	(0.158)&
Ti$_3$AlC$_2$	(0.158)&
Zr$_2$SnC	(0.158)\\
Ti$_2$SnC	(0.160)&
Ti$_3$SnC$_2$	(0.161)&
Ti$_2$AlC  	(0.164)\\
Nb$_5$AlC$_4$	(0.165)&
Ti$_3$IrC$_2$	(0.169)&
Hf$_2$AlN	(0.169)\\
Ti$_4$GaC$_3$	(0.173)&
Cr$_2$GeC	(0.173)&
Nb$_3$SiC$_2$	(0.175)\\
Nb$_4$AlC$_3$	(0.175)&
Ti$_2$GaC	(0.176)&
Ta$_5$AlC$_4$	(0.177)\\
Ti$_3$SiC$_2$	(0.181)&
Ti$_4$SiC$_3$	(0.181)&
Cr$_2$TiAlC$_2$	(0.184)\\
Nb$_2$GaC	(0.185)&
Nb$_2$AlC	(0.185)&
Ti$_3$GeC$_2$	(0.186)\\
Nb$_2$AsC	(0.186)&
V$_2$GeC	(0.186)&
V$_2$AsC	(0.186)\\
Ti$_4$GeC$_3$	(0.186)&
Ta$_2$GaC	(0.187)&
Ti$_2$SiC (0.188)\\
Cr$_2$GaN	(0.189)&
Ti$_2$GeC	(0.189)&
V$_3$SiC$_2$	(0.190)\\
Ti$_2$AlN	 (0.190)&
Mo$_2$GaC	(0.191)&
Ta$_2$AlC	(0.192)\\
Ti$_4$AlN$_3$	(0.193)&
Ti$_2$AsC	(0.195)&
Mo$_2$TiAlC$_2$ (0.196)\\
Cr$_2$GaC	(0.197)&
Ti$_2$GaN	(0.197)&
V$_3$AlC$_2$	(0.199)\\
Mo$_2$Ti$_2$AlC$_3$ (0.199)&
Mo$_2$ScAlC$_2$ (0.200)&
V$_4$AlC$_3$	(0.201)\\
V$_2$GaC	(0.202)&
V$_2$SiC	(0.205)&
V$_2$AlC	(0.205)\\
\hline
\end{tabular}
\label{tab:exfoliation}
\end{table}

 \subsection{Exfoliation energy analysis}
As we described above, the chemical exfoliation is a complicated dynamical process with subtle details. 
Hence, it is extremely difficult to simulate such processes computationally. However, the calculation 
of the static exfoliation energy would help us to screen MAX phases for the successful exfoliation 
process into 2D MXenes. 
Here, we evaluate the exfoliation energy $E_{\rm Exfoliation}$ for the 82 different MAX phases 
summarized in Table 2S. The definition of $E_{\rm Exfoliation}$ is given in the section of methods of calculations. 
As shown in Table~\ref{tab:experiment}, among the MAX phases experimentally exfoliated into MXenes, 
Zr$_3$AlC$_2$ (V$_2$AlC$_2$) shows the lowest (largest) $E_{\rm Exfoliation}$ (eV/area), 
$0.131$ ($0.205$) eV/\AA$^2$.
Therefore, we expect that the MAX phases with an exfoliation energy lower than $0.205$ eV/\AA$^2$ 
have a better chance to be exfoliated into MXenes. These MAX phases are listed in Table~\ref{tab:exfoliation}. 
The compounds listed in Table~\ref{tab:exfoliation} are sorted based on their exfoliation 
energies in ascending order: the smaller the exfoliation energy is, the better the chance for successful 
exfoliation becomes. 
Therefore, the table lists the compounds from the most promising candidate to the least promising one 
for successful exfoliation.
Figure~\ref{fig:exfoliation}(a) shows the total force constant FC$_{\rm A}$ for the A atoms 
versus the exfoliation energy. 
It is clearly observed that the total force constant FC$_{\rm A}$ is strongly correlated with the 
exfoliation energy: the MAX phases with smaller FC$_{\rm A}$ have a lower exfoliation energy.

Because of the correlation between the force constant FC$_{\rm A}$ and the exfoliation energy, 
one would immediately expect a similar correlation with the elastic constant C$_{33}$ along the $c$ direction. 
This is simply because the M$_1-$A bonds are almost parallel to the $c$ axis and thus 
C$_{33}$ should be related to the force constant FC$_{\rm A}$. 
Therefore, we calculate the elastic constants C$_{33}$ for all MAX phases. 
As shown in Fig.~\ref{fig:exfoliation}(b), we indeed find that 
C$_{33}$ is correlated with the exfoliation energies. 
This suggests that MAX phases with smaller (larger) C$_{33}$ are easier (more difficult) to be exfoliated 
into 2D MXenes. 
However, we note that C$_{33}$ should be used as a criterion for the exfoliation with care, 
because when the numbers of M and X layers are larger in the unit cell, 
as in M$_{n+1}$AX$_n$ ($n= 2-4$) MAX phases, 
the force constants for M and X atoms should also be relevant for C$_{33}$ and thus the correlation 
between C$_{33}$ and the exfoliation energy becomes weak. 
In any case, our finding here is important since C$_{33}$ is an experimentally measurable quantity.

 \begin{figure}[t]
\centering
  \includegraphics[scale=0.39]{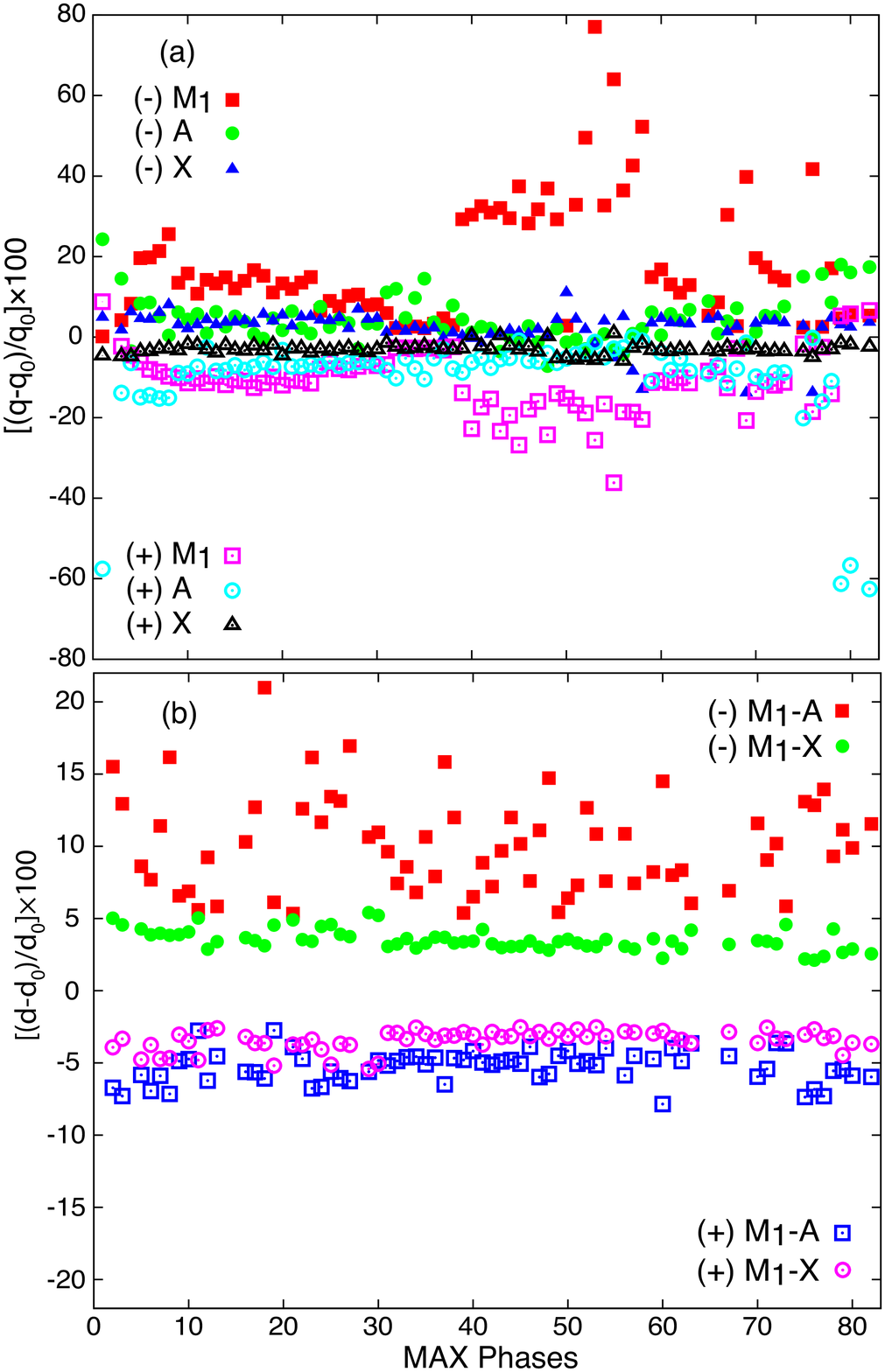}
  \caption{(a) The relative amount of charge acquired by each element of the 82 experimentally synthesized 
  MAX phases.  $q$ ($q_0$) represents the amount of charge on the M, A, and X atoms in the charged 
  (neutral) MAX phases. 
  (b) The relative change of length of the M$_1-$A and M$_1-$X bonds for the 82 experimentally synthesized 
  MAX phases. $d$ ($d_0$) denotes the bond distance in the charged (neutral) MAX phases. 
  Plus (minus) sign in parenthesis indicates the positively (negatively) charged MAX phases. 
  The numbers 1$-$82 in the horizontal axis represents the experimentally synthesized MAX phases, 
  listed in Table 1S.
  }
  \label{fig:chargeeffect.eps}
\end{figure}

 \subsection{Best candidates}

Combining all the analyses, we can finally conclude that the best candidates of MAX phases 
for the successful exfoliation into 2D MXenes are those MAX phases that are common in 
Tables~\ref{tab:FCA},~\ref{tab:FCX}, and~\ref{tab:exfoliation}, \textit{i.e.}, MAX phases with 
small total force constants of the A atoms, large total force constants of the M and X atoms, and 
low exfoliation energy. 
We predict that in addition to the 10 MAX phases that have 
already been exfoliated experimentally (listed in Table~\ref{tab:experiment}), the following 37 MAX phases are 
promising candidates for the successful exfoliation. 
Since the exfoliation energy is more directly related to the chemical 
exfoliation process than the other two criteria, we  prioritize this criterion to sort these 37 promising candidates 
from the smallest to the largest exfoliation energies.
Therefore, in the following list, we predict that the first (last) MAX phase is the most (least) promising candidate 
for successful exfoliation: 
1) Ti$_2$CdC, 2) Zr$_2$AlC, 3) Ti$_3$AuC$_2$, 4) Ti$_5$AlC$_4$, 5) Zr$_2$InC, 6) Hf$_2$AlC, 7) Ti$_2$GaC,
8) Ti$_4$GaC$_3$, 9) Hf$_2$InC, 10) Nb$_5$AlC$_4$, 11) Hf$_2$TlC, 12) Ti$_2$InC,
13) Ti$_2$TlC, 14) Nb$_2$GaC, 15) Hf$_2$PbC, 16) Ta$_5$AlC$_4$, 17) Ti$_3$SiC$_2$, 18) Ti$_4$SiC$_3$, 
19) Ti$_3$GeC$_2$, 20) Nb$_2$InC, 21) Ti$_2$GeC, 22) Ti$_4$GeC$_3$, 23) Hf$_2$SnC, 24) Mo$_2$GaC, 25) Ti$_2$SiC,
26) Ta$_2$GaC, 27) Cr$_2$GaN, 28) Ti$_3$IrC$_2$, 29) V$_2$GaC, 30) Ti$_2$InN, 31) Ta$_2$AlC$_2$, 32) Ti$_2$PbC, 
33) Ta$_2$AlC, 34) Cr$_2$GaC, 35) V$_4$AlC$_3$, 36) V$_3$AlC$_2$, and 37) Ti$_2$SnC.

\begin{figure*}[t]
\centering
  \includegraphics[scale=0.62]{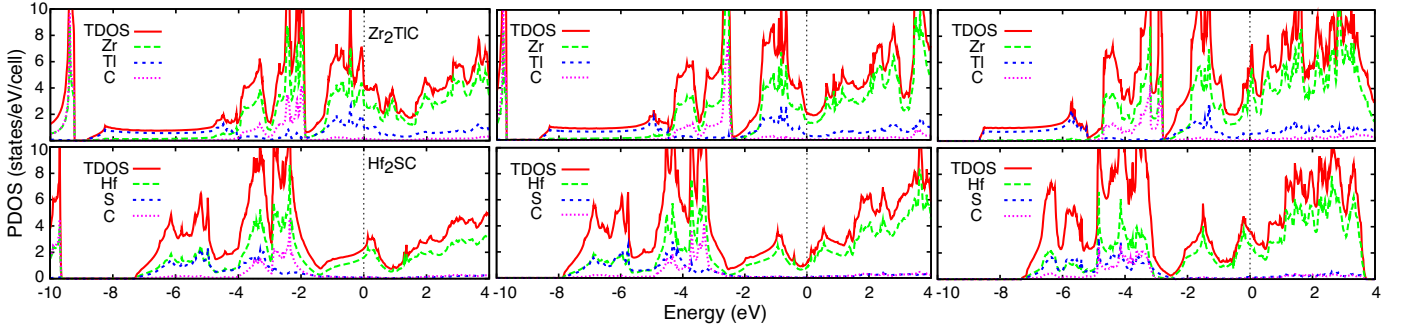}
  \caption{
  Projected density of states (PDOS) onto each atomic species (indicated in the figures) for positively, neutral, and negatively 
  charged (from left to right) Zr$_2$TlC and Hf$_2$SC (in top and bottom panels). 
  For comparison, the total density of states 
  (TDOS) is also shown. A vertical dotted line at zero energy indicates the Fermi energy.}
  \label{fig:pdos}
\end{figure*}

 \subsection{Effect of charging} 
 
In order to investigate the effects of electron (hole) injection to the MAX phases, here we add (remove) 
2, 3, 4, and 5 electrons per unit cell to (from) M$_2$AX, M$_3$AX$_2$, M$_4$AX$_3$, and 
M$_5$AX$_4$, respectively by changing the total number of valence electrons.
The number of injected charges is chosen such that the number of injected charges per atom in all the 
studied MAX phases is the same. 
There are two options to analyze the charge distribution. One is the integration of PDOS up to the Fermi level, in which the eigenstates are projected onto atomic orbitals that are defined in the pseudopotentials and thus do not form a complete basis. Therefore, the charge conservation is violated in a sense that the number of electrons on the atoms is not found correctly.
The second option, which satisfies the charge conservation, is the Bader charge analysis, based on a real-space grid.\cite{G.Henkelman2006,E.Sanville2007,W.Tang2009}
We adopt the second option for our charge analysis.

Figure~\ref{fig:chargeeffect.eps}(a) shows the relative number of electrons or holes received by the M, A, and X 
atoms in the charged MAX phases with respect to the neutral ones. 
Interestingly, in both positively and negatively charged MAX phases, most of the charges are 
received by the transition metals. 
That suggests that the states close to the Fermi energy must be transition-metal-centered. 
In order to explore this observation, we calculate 
the electronic structure of the charged and neutral MAX phases. 
Typical results for the projected density of states for charged and neutral Zr$_2$TlC and Hf$_2$SC 
are shown in Fig.~\ref{fig:pdos}.
More results are provided in the supporting information file for various 
MAX phases, 
including Zr$_2$TlN, Ti$_3$AuC$_2$, Nb$_4$AlC$_3$, Ti$_5$AlC$_4$, Mo$_2$ScAlC$_2$, 
and Mo$_2$Ti$_2$AlC$_3$. 
It is found that in all neutral MAX phases, the states near the Fermi energy are dominated by $d$ orbitals of 
the transition metals, as anticipated already. 
Hence, the electrons or holes injected into the MAX phases are mainly received by the transition metals. 
Among Zr$_2$TlC and Hf$_2$SC, Hf atoms in Hf$_2$SC receive a larger number of electrons or holes 
as compared with Zr atoms in Zr$_2$TlC.  
This is simply because in Hf$_2$SC the states below and above the Fermi energy are almost purely 
contributed by the $d$ 
orbitals of Hf, while in the case of Zr$_2$TlC, in addition to the $d$ orbitals of Zr, there is a contribution 
from the $d$ orbitals of Tl near the Fermi energy.

In order to investigate the effect of charge injection on the bonds, we calculate in Fig.~\ref{fig:chargeeffect.eps}(b) 
the relative change of the M$_1-$A and M$_1-$X bond lengths. It is found that in the positively (negatively) 
charged MAX phases, the bond lengths of both M$_1-$A and M$_1-$X bonds decrease (increase).  
By injecting electrons (holes), the number of electrons on the M, A, or X atoms increases (decreases), 
which results in a longer (shorter) bond length. 
Interestingly, the impact of the charge injection on the M$_1-$A bonds is generally more significant 
than that on the M$_1-$X bonds. The results for the M$_1-$A bond elongation with the electron 
injection are summarized in Table~\ref{tab:elongation}. 
When the transition metals receive a sufficient amount of electrons, 
the M$_1-$A bonds expand and may become completely broken. Indeed, we find that 
in the negatively charged Cr$_2$GeC, V$_2$GeC, Ti$_3$SnC$_2$, V$_4$AlC$_3$, and Cr$_2$TiAlC$_2$, 
the M$_1-$A bond length increases so large that the A atoms are completely detached from the transition metals. 
As an examples of electron-injected MAX phases, we can consider Li or Na intercalated MAX 
phases.\cite{J.Xu2017,A.T.Tesfaye2017} Our results imply that the electron donated by Li or Na atoms 
can make the intercalation easier.

\begin{table}
\caption{
List of MAX phases sorted in descending order (from top left to bottom right) by the relative bond elongation 
$\Delta d\, (= 100 \times\frac{d-d_0}{d_0})$ of the M$_1-$A bonds after receiving the same amount of 
electrons per atom. 
Here, $d$ ($d_0$) is the M$_1-$A bond distance in the negatively charged (neutral) MAX phases and 
only the MAX phases with $\Delta d > 10$ are listed. 
``Detached'' indicates that the A atoms are completely detached from the transition metals 
upon electron injection. 
The results for all studied MAX phases are provided in the supporting information file.  
}
\small
\begin{tabular}{ccc}
\hline
V$_2$GeC (detached) & 
Cr$_2$GeC (detached) &
Ti$_3$SnC$_2$ (detached) \\ 
V$_4$AlC$_3$ (detached) &
Cr$_2$TiAlC$_2$ (detached) &
 Ti$_2$PbC	(20.987) \\
V$_3$AlC$_2$	(18.286) &
V$_3$SiC$_2$	(16.976) &
V$_2$GaN	(16.950) \\
Sc$_2$TlC	(16.171) &
Ti$_2$TlC	(16.155) &
Nb$_2$SC	(15.845) \\
Cr$_2$GaC	(15.515) &
Ta$_5$AlC$_4$ (15.439)&
Cr$_2$AlC	(14.953) \\
Zr$_2$TlN	(14.729 &
Ti$_3$AuC$_2$	(14.506) &
Nb$_5$AlC$_4$	(13.942) \\
V$_2$AsC	(13.446) &
V$_2$GaC	(13.146) &
Nb$_4$AlC$_3$	(13.099) \\
Cr$_2$GaN	(12.949) &
Ta$_4$AlC$_3$	(12.851) &
Ti$_2$InN	(12.712) \\
Hf$_2$PbC	(12.674)&
Ti$_2$SnC	(12.601) &
Zr$_2$PbC	(12.002) \\
Nb$_2$SnC	(12.000) &
V$_2$AlC	(11.672 &
Ti$_2$SC	(11.593) \\
Ti$_4$AlN$_3$	(11.592) &
Mo$_2$Ti$_2$AlC$_3$	(11.549) &
Sc$_2$InC	(11.411) \\
Mo$_2$ScAlC$_2$	(11.153) &
Zr$_2$TlC	(11.111) &
V$_2$SiC	(10.971) \\
Hf$_2$TlC	(10.870) &
Hf$_2$SC	(10.852) &
Nb$_2$InC	(10.650) \\
V$_2$PC	(10.643) &
Ti$_2$InC	(10.311) &
Ti$_4$GeC$_3$	(10.185) \\
 \hline
\end{tabular}
\label{tab:elongation}
\end{table}

 \begin{figure}[t]
\centering
  \includegraphics[scale=0.33]{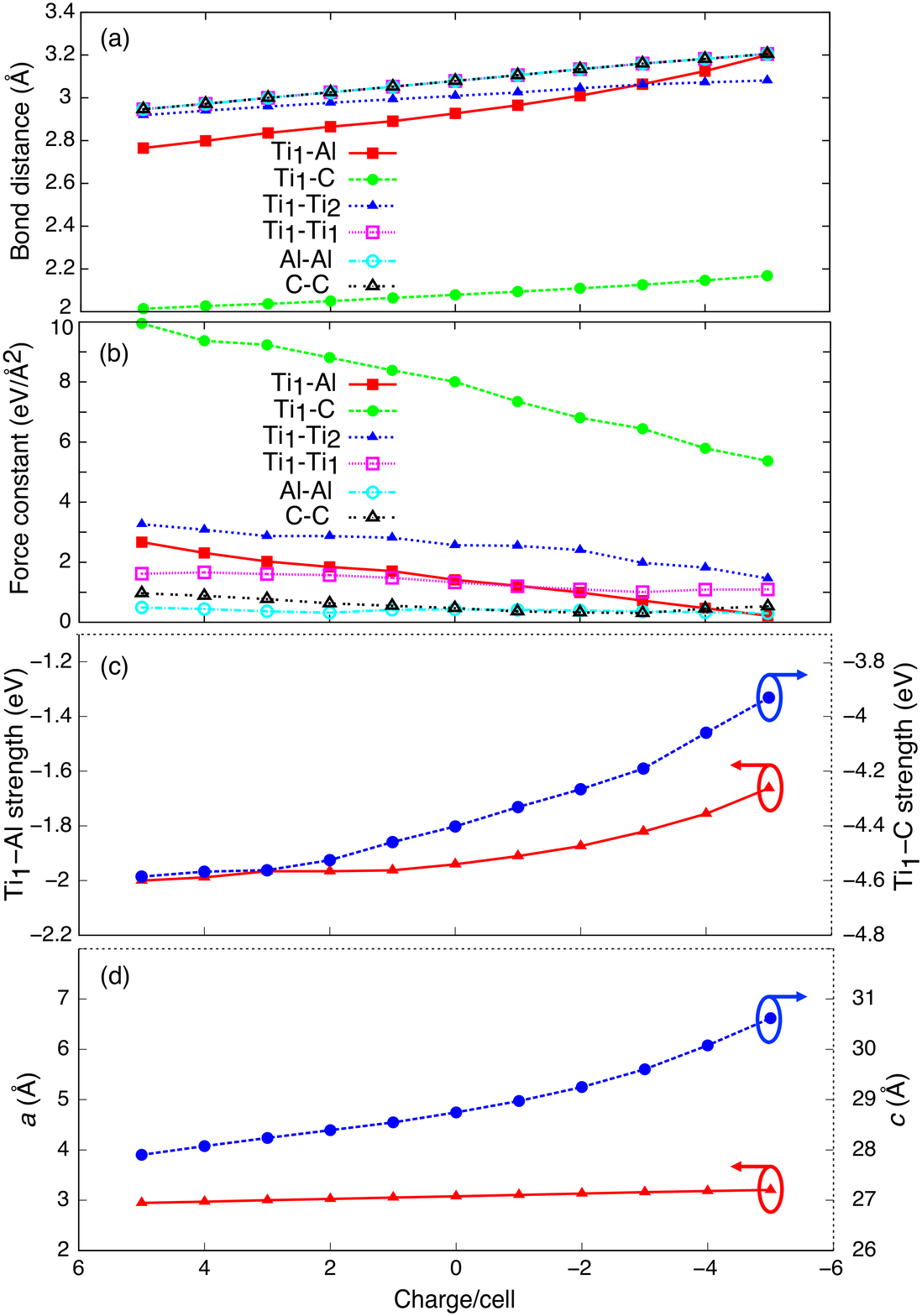}
  \caption{
  (a) Bond distances, (b) force constants, (c) bond strength, and (d) lattice parameters of Ti$_5$AlC$_4$ 
  when electrons or holes are injected. The horizontal axis represents the number of electrons (negative) or holes 
  (positive) injected per unit cell and zero corresponds to the neutral system. 
  }
  \label{fig:ti}
\end{figure}

In order to understand clearly the effect of charge injection on the structural properties, 
thicker MAX phases such as Ti$_5$AlC$_4$ and Nb$_5$AlC$_4$ are preferable 
than thinner ones. This is because the thinner MAX phases 
become unstable with only a few electrons injected. 
Although Ti$_5$AlC$_4$ and Nb$_5$AlC$_4$ 
have not been synthesized as phase-pure ternary compounds yet, their solid solutions 
(Ti$_{0.5}$Nb$_{0.5}$)$_5$AlC$_4$ already exist experimentally.\cite{L.Zheng2010} 
Therefore, we examine these two MAX phases Ti$_5$AlC$_4$ and Nb$_5$AlC$_4$ here.
Figure~\ref{fig:ti} summarizes the results for the bond lengths 
and the corresponding force constants $F_{ij}$, as well as the bond strength and the 
lattice constants, when 1$-$5 electrons or holes per unit cell are injected into Ti$_5$AlC$_4$.  
As shown in Fig.~\ref{fig:ti}(a), upon the electron (hole) injection, all bond lengths tend to increase 
(decrease), and the change of the bond length is larger when the amount of injected charge is larger. 
It is also clearly observed that the Ti$_1-$Al bond is most sensitive to the charge injection. 
Moreover, as shown in Fig.~\ref{fig:ti}(b), 
the force constants increases (decreases) with the hole (electron) injection. 
This occurs because the positive charge injected on either of atoms in the bond would increase the 
electronegativity and decrease the bond length, resulting in the increase of the ionic contribution 
to the force constant [$\sim (\chi_X\chi_Y/r^2)^{3/4}$]. 
Notice also in Fig.~\ref{fig:ti}(b) that after injecting a certain amount of electrons, the force constant 
of the Ti$_1-$Al bond becomes very small. This implies that if a larger amount of electrons is injected, 
the A atom (\textit{i.e.}, Al) becomes detached from the transition metal Ti.

Since the charge injection alters the bond lengths, the bond strength, 
\textit{i.e.}, the orbital interaction in the bond, should also be affected. In order to analyze this effect, 
we also calculate in Fig.~\ref{fig:ti}(c) the ICOHP up to the Fermi energy for the Ti$_1-$C and Ti$_1-$Al 
bonds of Ti$_5$AlC$_4$ when electrons or holes are injected. The additional analysis is also provided 
in the supporting information file. 
As shown in Fig.~\ref{fig:ti}(c), regardless of the type and amount of the injected charge, the Ti$_1-$C bonds 
are always stronger than the Ti$_1-$Al bonds. 
Upon the electron (hole) injection, the Ti$_1-$C and Ti$_1-$Al bond lengths increase (decrease), and thus 
their atomic interactions become weaker (stronger). This is consistent with the force constants shown 
in Fig.~\ref{fig:ti}(b). The analysis of the projected phonon density of states for charged 
Ti$_5$AlC$_4$ also reveals these trends (see the supporting information file). 
Upon the electron injection, all phonon frequencies decrease and, particularly, the phonon modes projected 
onto the Al 
atom become localized at very low phonon frequencies. 
By injecting a large amount of electrons, the phonon frequencies of the Ti$_1-$Al bond
become almost zero, indicating that the Al atom is almost detached 
from the Ti atom. 
It should be noted that the phonon calculations indicate the structural instability 
of positively charged MAX phases because we find that Ti$_5$AlC$_4$ becomes unstable 
after removing six electrons. This is due to depletion of some of the bonding states. 
Therefore, upon removing many electrons, the system becomes unstable and might undergo 
a structural phase transition.

We also investigate in Fig.~\ref{fig:ti}(d) the effect of charge injection on the $a$ and $c$ lattice parameters of 
Ti$_5$AlC$_4$. 
As expected, the strongest effect occurs in the $c$ lattice constant because the Ti$_1-$Al bonds are 
oriented almost parallel to the $c$ axis. 
In particular, the electron injection induces large swelling along the $c$ axis in Ti$_5$AlC$_4$ and 
also in other MAX phases in general. This implies that the exfoliation 
process becomes easier for the negatively charged MAX phases than 
for the neutral case simply because the M$_1-$A bonds become longer and weaker upon receiving electrons. 
The exfoliation of charged MAX phases might be possible experimentally through charge-controlled 
electrochemical swelling techniques.\cite{J.C.P.Gabriel2001,F.Kaasik2013,A.F.Nieves1999,J.K.Riley2014,S.Rosenfeldt2016,D.A.Laird2007,K.K.Mohan1997,H.Suquet1976,A.Fitch1995,R.Zahn2014,M.C.Pazos2016}
The volume swelling can also occur using irradiation methods 
by controlling the density of intrinsic point defects.\cite{J.M.Pruneda2004}

Finally, we should note that the atomic position relaxation is essential  
to correctly capture the trends found above for the charged MAX phases. 
We have carried out the additional systematic calculations without structural relaxation by using a rigid-band model and 
the results are shown in the supporting materials files. In particular, we emphasize that 
the weakening of the bond strength with the electron injection found in Fig.~\ref{fig:ti}(c) would not 
occur without the atomic position relaxation (as seen in the rigid-band approach), 
signifying that this is an elastic, not an electronic effect.

We should also remark on the calculations for charged MAX phases.
Although the GGA/PBE method can predict the physics of charged MAX phases, the results 
for the cases of many electrons or holes injected might be biased by the method because the electrons are treated as being very delocalized. 
This might be more severe in the case of positively charged MAX phases because one can naively expect that 
when the transition metals become significantly positively charged after removing a certain number of electrons, 
the enhanced Coulomb repulsions between the transition metal ions cause the lattice expansion. 
However, we have observed this expansion in our calculations for negatively charged MAX phases, but not for 
positively charged ones. 
This asymmetrical behavior is related to the way that positive and negative charges are distributed in the cell. 
In both negatively and positively charged cases, there exists a charge neutralizing background and thus 
the only difference is the extended nature of extra electrons versus the localized nature of extra holes 
around atomic cores. The repulsion is clearly larger in the case of extended electrons that are closer to 
each other (on the average) as compared with positive ions, the distance of which is never smaller than 
the bond lengths. Therefore, we expect that a 
suitable self-interaction corrected functional would describe the localizations of electrons more accurately. 
However, in general, we expect a larger repulsion in negatively charged systems than in positively charged ones.

\section{CONCLUSION}

MAX phases can be a great source for the synthesis of novel 2D materials with exceptional properties. 
So far, around 82 different crystalline and numerous alloy MAX phases have been synthesized.
Here, we have provided an insight into the exfoliation possibility of various crystalline MAX phases 
by examining the strength of bonds through 
force constant calculations, COHP analysis, and the bond order analysis. 
Our systematic analyses show that in all MAX phases, the M$_1-$X bonds are stiffer and stronger 
than the M$_1-$A bonds. 
The large stiffness of the M$_1-$X bonds is attributed to the greater orbital mixing and higher ionicity of the 
M$_1-$X bonds than those of the M$_1-$A bonds. 
This is also consistent with the bond order analysis, showing that the M$_1-$X bonds possess 
higher total number of bonds than the M$_1-$A bonds. 
The total force constant for the A atoms in MAX phases is found to be linearly correlated 
with the chemical exfoliation 
energy, and therefore the force constant can be used to investigate the bond strength and the exfoliation 
likelihood in MAX phases. 
The elastic constant C$_{33}$ along the $c$ direction supports this general trend: we have 
found that C$_{33}$ is also linearly correlated with the exfoliation energy. 
This implies that the MAX phases with large C$_{33}$ are difficult to be exfoliated into 2D MXenes. 
This argument for C$_{33}$ is best applied to the M$_2$AX MAX phases where 
the number of A layers in the unit cell is comparable with the number of M and X layers.

We have found that except for MAX phases with the A element of S or P, many of the MAX phases 
are promising candidates for exfoliation into 2D MXenes. 
Our comprehensive analyses predict that the following MAX phases have a better chance 
to be exfoliated successfully into 2D MXenes:
Ti$_2$CdC, Zr$_2$AlC, Ti$_3$AuC$_2$, Ti$_5$AlC$_4$, Zr$_2$InC, Hf$_2$AlC, Ti$_2$GaC,
Ti$_4$GaC$_3$, Hf$_2$InC, Nb$_5$AlC$_4$, Hf$_2$TlC, Ti$_2$InC,
Ti$_2$TlC, Nb$_2$GaC, Hf$_2$PbC, Ta$_5$AlC$_4$, Ti$_3$SiC$_2$, Ti$_4$SiC$_3$,
Ti$_3$GeC$_2$, Nb$_2$InC, Ti$_2$GeC, Ti$_4$GeC$_3$, Hf$_2$SnC, Mo$_2$GaC, Ti$_2$SiC,
Ta$_2$GaC, Cr$_2$GaN, Ti$_3$IrC$_2$, V$_2$GaC, Ti$_2$InN, Ta$_2$AlC$_2$, Ti$_2$PbC, 
Ta$_2$AlC, Cr$_2$GaC, V$_4$AlC$_3$, V$_3$AlC$_2$, and Ti$_2$SnC. 
Therefore, we expect that various 2D MXenes can be synthesized, including 
Ti$_2$C, Ti$_3$C$_2$, Ti$_4$C$_3$, Ti$_5$C$_4$, 
Ti$_2$N, Zr$_2$C, Hf$_2$C, V$_2$C, V$_3$C$_2$, V$_4$C$_3$, Nb$_2$C, Nb$_5$C$_4$, Ta$_2$C, Ta$_5$C$_4$
Cr$_2$C, Cr$_2$N, and Mo$_2$C.

Moreover, we have analyzed the charged MAX phases that exhibit unique electronic properties. 
Since the states near the Fermi energy are dominated by $d$ orbitals of the transition metals, 
the injected charges are mainly received by the transition metals. 
We have shown that the charge injection affects the M$_1-$A bonds most significantly. 
Upon receiving electrons, the M$_1-$A bonds are elongated, and thus the structure is swollen along the 
$c$ axis, 
which leads to the decrease of the force constants, 
thereby facilitating the exfoliation. 

\section*{Conflicts of interest}
There are no conflicts to declare.

\section*{Acknowledgements}
We are grateful to Dr. Rahul Maitra for fruitful discussion. M.K. and A.R. are also grateful to RIKEN Advanced 
Center for Computing and Communication (ACCC) for the allocation of computational resource of the RIKEN 
supercomputer system (HOKUSAI GreatWave). Part of the calculations were also performed on Numerical 
Materials Simulator at National Institute for Materials Science (NIMS). M.K. gratefully acknowledges the support by Grant-in-Aid for Scientific Research 
(No. 17K14804) from MEXT Japan. D.B. and R.D. gratefully acknowledge the IT Center 
of RWTH Aachen University for providing computational resources and time.

\end{document}